\documentclass[a4paper,fleqn,usenatbib]{mnras}  
\usepackage{newtxtext,newtxmath}
\usepackage[T1]{fontenc}
\usepackage{ae,aecompl}
\usepackage{graphics,graphicx}
\usepackage{amsmath}	

\def\JKCS{JKCS\,041 }        
\def\IDCS{IDCS J1426.5+3508 }

\title[\JKCS ICM assembly]{
Witnessing the intracluster medium assembly at the cosmic noon in \JKCS
}
\author[S. Andreon et al.]{S. Andreon,$^{1}$\thanks{email:stefano.andreon@inaf.it} 
C. Romero$^{2,3,4}$, 
H. Aussel$^5$,
T. Bhandarkar$^4$, 
M. Devlin$^4$,
S. Dicker$^4$, 
\and
B. Ladjelate$^6$
I. Lowe$^{7,4}$,
B. Mason$^8$,
T. Mroczkowski$^9$,
A. Raichoor$^{10}$,
C. Sarazin$^{11}$,
\and
G. Trinchieri$^1$
\and  \\
$^1$ INAF--Osservatorio Astronomico di Brera, via Brera 28, 20121, Milano, Italy\\
$^2$ Center for Astrophysics $\vert$ Harvard \& Smithsonian, 60 Garden Street, Cambridge, MA 02138, USA \\
$^3$ Green Bank Observatory, 155 Observatory Road, Green Bank, WV 24944, USA \\
$^4$ Department of Physics and Astronomy, University of Pennsylvania, 209 South 33rd Street, Philadelphia, PA, 19104, USA \\
$^5$ AIM, CEA, CNRS, Universit\'e Paris-Saclay, Universit\'e Paris Diderot, Sorbonne Paris Cit\'e, 91191 Gif-sur-Yvette, France \\
$^6$ Instituto de Radioastronomia Milimetrica (IRAM), Granada, Spain \\
$^7$ Department of Astronomy, University of Arizona, Tucson, AZ, 85721, USA \\
$^8$ National Radio Astronomy Observatory, 520 Edgemont Road, Charlottesville, VA 22903, USA \\
$^9$ European Southern Observatory, Karl-Schwarzshild-Str. 2, D-85748 Garching b. M\"{u}nchen, Germany \\
$^{10}$ Lawrence Berkeley National Laboratory, 1 Cyclotron Road, Berkeley, CA 94720, USA\\
$^{11}$ Department of Astronomy, University of Virginia, P.O. Box 400325, Charlottesville, VA 22904, USA \\
}
\date{Accepted ... Received ..; in original form ..}
\pubyear{2020}

\begin{document}
\label{firstpage}
\pagerange{\pageref{firstpage}--\pageref{lastpage}}
\maketitle

\begin{abstract}
In this work we study the intracluster medium of a galaxy cluster at the cosmic noon: \JKCS at $z=1.803$.
A 28h long Sunyaev-Zel'dovich (SZ) observation using MUSTANG-2 allows 
us to detect \JKCS, even if 
intrinsically extremely faint compared to other SZ-detected clusters.
We found that the SZ peak is offset from the X-ray center by about 220 kpc in the direction of the brightest cluster galaxy, which we interpret as due to the cluster being observed just after first passage of a major merger. 
\JKCS  has a low central pressure and a low Compton $Y$
compared to local clusters selected by their intracluster medium (ICM), likely because the cluster is
still in the process of assembly but also in part because of a hard-to-quantify bias in current local ICM-selected samples. 
\JKCS  
has a $0.5$ dex fainter $Y$ signal than another less massive $z\sim1.8$
cluster, exemplifying how much different weak-lensing mass and SZ mass can be at high redshift.
The observations we present provide us with the measurement 
of the most distant resolved pressure profile of a galaxy cluster. Comparison with a library of plausibly descendants shows that \JKCS pressure profile will likely increase by about 0.7 dex in the next 10 Gyr at all radii.
\end{abstract}

\maketitle

\section{Introduction}

The redshift around two is a key epoch for galaxies and clusters:  the cosmic star formation rate and the black
hole growth peak at about this redshift (Madau \& Dickinson 2014) and  
the first massive clusters emerge from the cosmic web (Arnaud et al. 2009). At this epoch of high star formation rates, AGN
and merger activity, 
the intracluster medium (ICM) is heavily shaped by accretion shocks, heating from stellar winds associated to star formation
of member galaxies, and mechanical heating and turbulence generated by AGNs (McNamara et al. 2005). A characterization
of the ICM of clusters at this epoch would therefore offer a measurement of the efficiency of these feedback mechanisms
and of the gas circulation in these initial stages of the cluster formation.  However, in spite of extensive 
searches, in the last ten years we were unable to characterize the ICM of any 
cluster at $z>2$, the most distant characterized clusters being XLSSC122 (Willis et al. 2020) at $z=1.98$,
\JKCS (Andreon et al. 2009, Newman et al. 2014, Andreon et al. 2014) at $z=1.803$ and \IDCS (Stanford et al. 2012, 
Andreon et al. 2021) at $z=1.75$. At $z>2$, ICM detections are still quite uncertain;  
the ICM detection of a $z=2.5$ cluster by Wang et al. (2016) has been 
questioned by Champagne et al. (2021). The $z=2.16$ Spiderweb structure is 
dominated by the  wings of PSF of the central bright AGN and by inverse-Compton scattering on the jet emission (Tozzi et al. 2022). 
The latter components outshine the tentative ICM detection at energies larger than about 1 keV (Tozzi et al. 2022), while the available Chandra
data below 0.8 keV are noisy at best because of the contaminant on the optical blocking filter. The complexity of this
handful of arcsec wide sky region, with two bright AGNs, two bright jets and at best a very weak ICM emission, has been pointed out by the SZ analysis of Di Mascolo et al. (2023).
The properties of the ICM at this key epoch are therefore poorly known because of the
difficulty of acquiring spatially resolved data adequate for the purpose. XLSSC122 and \JKCS have X-ray radial
surface brightness profiles (Mantz et al. 2018, Andreon et al. 2009). \IDCS instead, thanks to spatially resolved
Sunyaev-Zel'dovich (SZ) and X-ray  data, has all its thermodynamic profiles determined. This allowed Andreon et al. (2021) 
to find a sizeable gas, heat, and entropy transfer at $r>300$ kpc in \IDCS, while the cluster central region has likely reached the final stage. Of course, with just
one example studied in detail, it is impossible to say whether the inside-out ICM assembly seen in \IDCS is a general feature of the whole
cluster population at those redshifts.   
As mentioned, $z\sim2$ is an epoch of intense merger activity. The ICM of several  major mergers at low to intermediate
redshifts have been studied in detail (e.g., Cl J0152-1357 and Bullet clusters, Massardi et al. 2010; Menanteau et al. 2012), 
but adequate data to investigate the ICM properties of one such object at $z\sim2$ are still lacking.   At this redshift, the fraction of disturbed clusters is expected to be high, making this investigation even more pressing.

In this paper we report very deep
Sunyaev-Zel'dovich (SZ) observations of \JKCS  sufficient to measure the radial profile of the cluster pressure
and therefore to study its intracluster medium.
\JKCS is an intermediate-mass cluster (a few $10^{14}$ M$_\odot$, Andreon et al. 2014) 
at $z=1.803$ (Newman et al. 2014; Andreon et al. 2014)
with hot gas detected by Chandra (Andreon et al. 2009).
With 19 spectroscopic members, a well defined red sequence,
and an extensive multi-wavelength coverage, this system is perhaps 
the most studied cluster in the high redshift Universe; for example it is the only cluster so far
with measurements of velocity dispersion of their individual passive galaxies (Prichard et al. 2017). Mostly
based on the dynamics of the cluster galaxies, \JKCS seems is suspected to have substructure and
to be a merger in progress (Prichard et al. 2017). Unlike other clusters, \JKCS is not ICM-selected (was discovered
as galaxy overdensity and later follow-up, Andreon et al. 2009) 
and therefore does not share the common bias of ICM-selected samples that have
clusters with large signals over-represented (and clusters with low signals underrepresented or absent), see e.g.
Andreon et al. (2011).

Throughout this paper, we assume $\Omega_M=0.3$, $\Omega_\Lambda=0.7$, 
and $H_0=70$ km s$^{-1}$ Mpc$^{-1}$. 
Results of stochastic computations are given
in the form $x\pm y$, where $x$ and $y$ are 
the posterior mean and standard deviation. The latter also
corresponds to 68\% uncertainties because we only summarize
posteriors close to Gaussian in this way. All logarithms are in base 10.

\begin{figure*}
\centerline{
\includegraphics[width=15truecm]{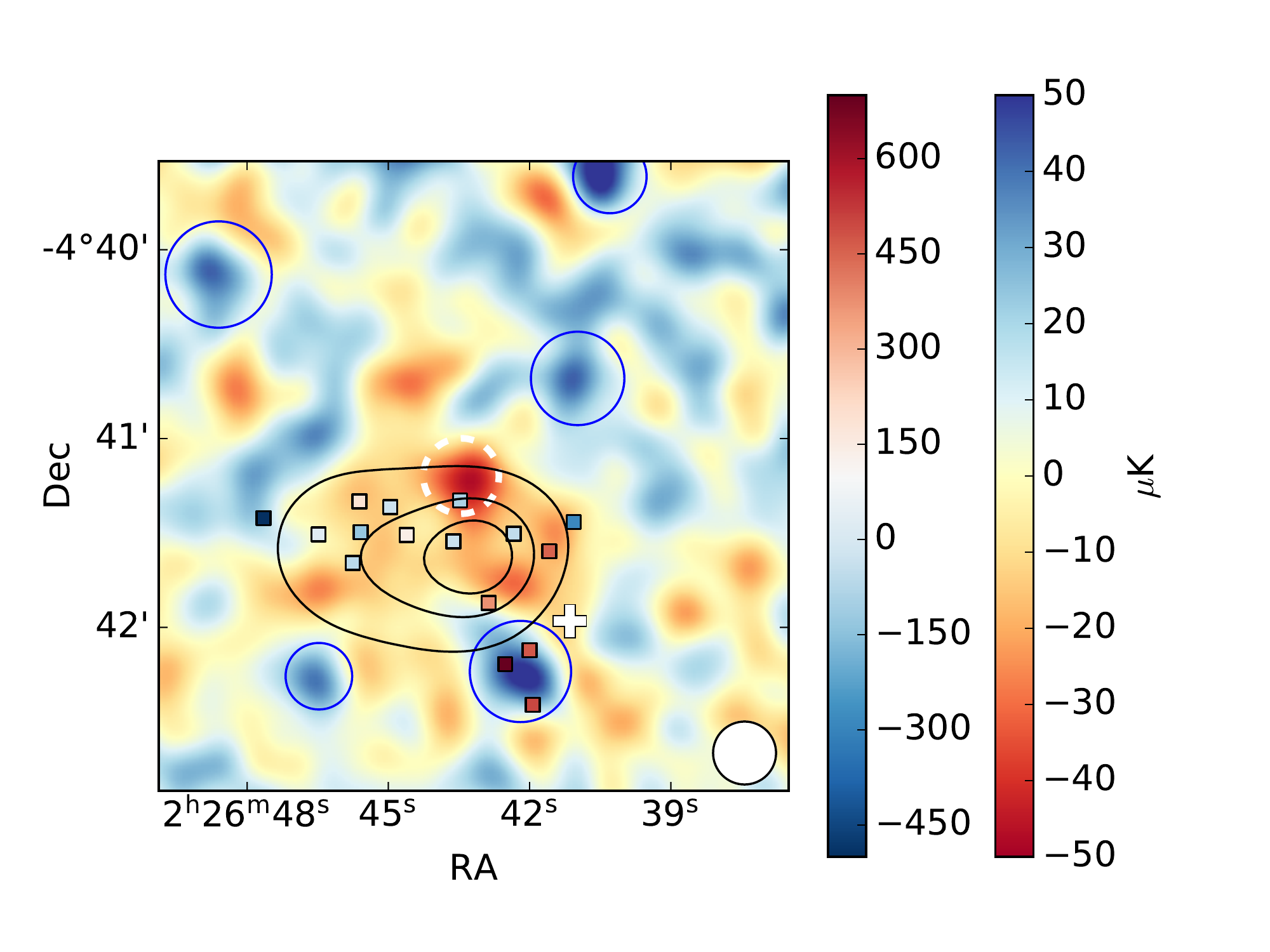}}
\caption[h]{Multi-wavelength view of \JKCS. 
The image is the 
MUSTANG-2 SZ map smoothed with a Gaussian with $\sigma=6$ arcsec. Large blue circles indicate the regions flagged in
our analysis because of the presence of point sources. The MUSTANG-2 beam is indicated in the bottom right. The SZ peak is indicated by a dashed white circle of 12 arcsec radius 
and the center
of a $z=1.13$ group is indicated by a plus sign. Black contours are X-ray isocontours of the 
adaptively-smoothed, optimally-extracted Chandra image. 
Spectroscopic confirmed cluster members are indicated by squares color-coded by relative 
velocity (inner color bar, in km s$^{-1}$). 
The SZ and X-ray peaks are offset by about 26 arcsec, in a nearly North-South direction.
Galaxies in the South have high positive velocities. 
The cluster brightest cluster galaxy is the square inside the white dashed circle. 
}
\label{fig:view}
\end{figure*}

\section{Observations, data reduction and analysis}

\subsection{Weak lensing}
\label{sec:wl}

Kim et al. (2023) performed
a weak lensing analysis of \JKCS based on WFC3 imaging data and photometric redshifts
presented in Newman et al. (2014). The authors found a mass of $M_{200}=4.7\pm1.5 \ 10^{14} M_{\odot}$.
The analysis makes a two-dimensional fit using Navarro, Frenk \& White (1997) profiles with the Diemer \& Joyce (2019) mass-concentration relation. Two mass components were jointly fitted, 
the \JKCS cluster itself and a spectroscopic confirmed $z=1.13$ group about 1 arcmin
south-west of the X-ray center (the plus sign in Fig.~\ref{fig:view}, Newman et al. 2014). 

A similar analysis for the other cluster  often mentioned in this work, \IDCS at $z=1.75$ 
returns a lower mass of $M_{200}=2.2^{+1.1}_{-0.7} \ 10^{14} M_{\odot}$ (Jee et al. 2017), in
agreement with its lower richness (Andreon 2013).
The $M_{500}$ masses of these two clusters, used here, have been derived from $M_{200}$ 
following Ragagnin et al. (2021).

\subsection{Galaxy dynamics}
\label{sec:dynamics}

Using the K-band Multi-Object Spectrograph at VLT,
Prichard et al. (2017) determined precise ($\lesssim 100$ km/s) redshifts of 16 \JKCS galaxies that were
known to be members using Hubble Space Telescope redshifts (with $\sim 500$ km/s precision) in Newman et al. (2014). 
These are shown in Fig.~\ref{fig:view} as 
squares color-coded by relative velocity. The spatial distribution of the relative
velocities (note that the red squares are all in the
cluster South) suggest that \JKCS is composed by two merging sub-clusters (Prichard et al. 2017).

\subsection{X-ray}
\label{sec:X}

\JKCS was observed with \emph{Chandra} ACIS-S for 75 ks; data reduction and analysis are described in 
Andreon et al. (2009). Here we use the same data to derive
an image to be used as comparison to the SZ image.

To optimize the imaging of the cluster, we start from 
Scharf (2002), who discusses the optimal energy range to be used 
to image galaxy clusters at $z<1$ with maximal signal to noise when photons cannot be weighted but can
only be taken or discarded 
depending on their energy. 
We improve upon his work eliminating the redshift limit and 
weighting photons with a weight given by the information content of each (infinitesimal) energy bin.
The weight is computed as the ratio between
the cluster and the background spectra. The cluster is modelled with an APEC spectrum with the  temperature measured for \JKCS(from Andreon et al. 2009) and its redshift (from Newman et al. 2014)
to enforce regularity. The background is measured from the photons falling in the same chip 
but outside the cluster
solid angle. Compared with the common
[0.5,2] keV choice, the optimal weight resemble a Maxwell distribution, with a peak at about 1 keV, a FWHM of 0.9 keV,
and a tail a higher energies. The decrease at 
$\sim 1.7$ keV is due to the presence of a background line there (as already remarked by Scharf 2002),  
while the decrease at lower energies is due to the increased background.  After weighing and getting rid
of point sources (interpolating over the tiny regions occupied by them), we apply an adaptive filter to the data
using CIAO csmooth (based on Ebeling et al. 2006) to  keep only features with $S/N>3$,
as already done in Andreon et al. (2009).  Isocontours of the image so obtained are  shown
in 
Fig.~\ref{fig:view}.

\subsection{SZ data}
\label{sec:SZ}

\JKCS was observed for 28h (overhead included)  with MUSTANG-2 
between December 2018 and January 2021.
MUSTANG-2 is a 215-element array
of feedhorn-coupled TES bolometers (Dicker et al. 2014) at the
100m Green Bank telescope (GBT) working at 90 GHz. When mounted on the GBT, MUSTANG-2
has about 10" FWHM resolution and an instantaneous field of view of 4.5 arcmin.

\JKCS was observed with Lissajous daisy scans, typically with a 2.5 arcmin radius. The data
were calibrated using preferentially Solar systems objects, and also 
ALMA calibrators (Fomalont et al. 2014, van Kempen et al. 2014).
Our SZ maps are calibrated to Rayleigh-Jeans brightness temperature ($K_{\text{RJ}}$), adopted 
throughout this paper.
The final map has a RMS noise of 23 $\mu K$ when smoothed to the beam
resolution, within the central 2 arcmin radius. 
A beam profile specific to
this cluster observations is created by stacking all secondary (point-source)
calibrators interspersed during these observations (see Ginsburg et al. 2020). 
The beam FWHM of the final map is 9.5 arcsec.

\begin{figure}
\centerline{\includegraphics[width=9truecm]{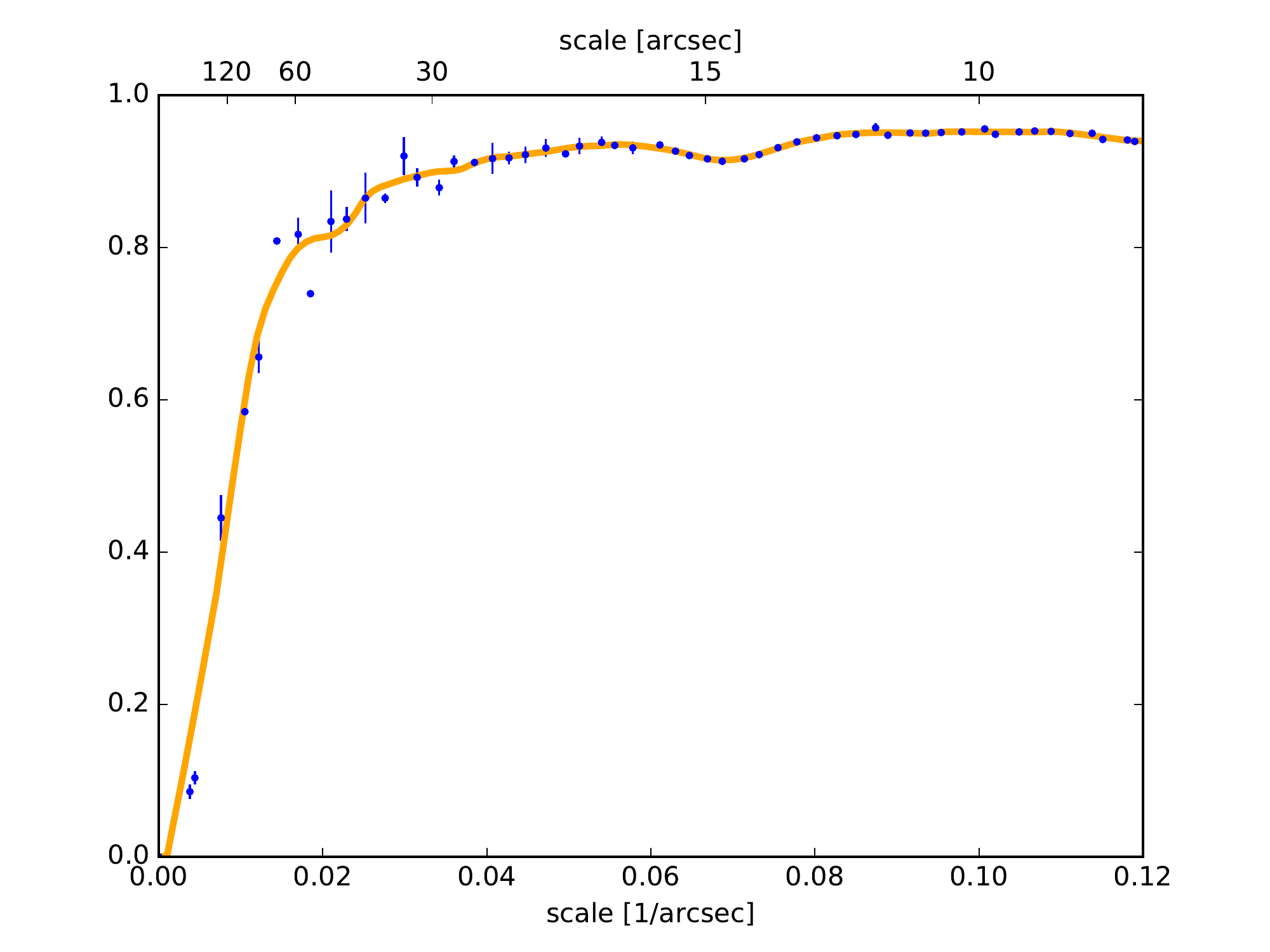}}
\caption[h]{\JKCS MUSTANG-2 transfer function. Signal at scales smaller than 1 arcmin is, by and large,
preserved by the MIDAS data reduction but is increasingly reduced with increasing scales.
}
\label{fig:tf}
\end{figure}

The data were reduced following the traditional approach (named MIDAS in
Romero et al. 2020, to which we refer for details): time ordered data from unresponsive detectors or
from bad scans and glitches are flagged, remaining ones are filtered to subtract
atmospheric and electronic background. This filtering also removes
some (cluster) signal, which needs to be accounted for during the analysis
using the transfer function, shown in Fig.~\ref{fig:tf}. 
The transfer function is larger than 0.8 at scales smaller than 1 arcmin, then drops to lower values.
Since atmospheric conditions were less stable during the observations of  \JKCS than of \IDCS,  
a more aggressive filtering was needed for \JKCS data. This resulted in a transfer function that 
drops to low values at smaller scales (compared to the one of \IDCS, shown in Andreon et al. 2021).

The resulting map is shown in  the left panel of Fig.~\ref{fig:view}. The SZ peak is at  (RA, Dec)=(2:26:43.2,-4:41:13). A few point sources are visible 
(dark blue in the Figure surrounded by blue circles) and pixels contaminated by them
are flagged and ignored during further cluster analysis.

For completeness, we report that 
\JKCS was also observed for 17h with the NIKA2 camera (Adam et al. 2018) at 150 and 250 GHz at the IRAM 30m telescope. 
The cluster SZ signal is not obvious in the preliminary reduction of
this shorter exposure taken with a smaller telescope, although 
there is a (negative) blip in the NIKA2 150 GHz map at the location of the MUSTANG-2 (negative) peak. 
We defer a full analysis of the properly reduced NIKA2 data to a later paper.

\subsection{The SZ-X offset}

The X-ray peak is at (RA,Dec)=(2:26:43.1,-4:41:36), about 26 arcsec (about 220 kpc) 
South of the SZ one derived in
Sec.~\ref{sec:SZ}
A spatial offset between
the X-ray and SZ centers is possible because of the different dependence of these two observables on
electron density and temperature. This has already been  seen for 
example in the merging clusters
Cl J0152-1357 (Massardi et al. 2010) and Bullet (Menanteau et al. 2012).
According to hydrodynamic simulations in Zhang et al. (2014), large
offsets are produced just after the pericenric passage 
during near head-on major mergers of clusters with $M> 10^{14} M_\odot$. 
Therefore, we interpret the SZ-X offset as due to the cluster being observed just after a major merger, 
as a part of the emergence of the first massive
clusters from the  cosmic web at early times. 
The spatial SZ-X offset corroborates
the suggestion by Prichard et al. (2017) that \JKCS still shows traces in the galaxy dynamics
of the two sub-clusters that formed it; see Fig.\ref{fig:view} with large positive velocities (red) points regrouped in the cluster south. In \JKCS, the SZ-X offset is in the North-South
direction, as the two groups identified by Prichard et al. (2017) (the red points).

Judging from
Zhang et al. (2014) simulations, the main cluster is coincident with the SZ peak while the
colliding cluster is South of it and came from North with $\Delta v \approx 500$ km/s and is seen $\approx 0.5$ Gyr 
after the pericentric after a nearly head-on collision. 
This would produce an X-ray peak about 250 kpc
South of the SZ peak, as observed in \JKCS. In the simulations, the two sub-clusters have a mass
of 1-2 $10^{14}$ M$_\odot$ (that are cluster masses at $z=1.8$) and their summed mass is
consistent with the observed
\JKCS weak lensing mass. Tailored (and more precise)  constraints on the \JKCS colliding
sub-clusters (e.g. gas fraction) and merger parameters (e.g. relative velocity and projection effects) accounting for the observational setup (e.g., SZ transfer function) and constrained by measured quantities (mass, Compton-Y, etc.)
are presented in Felix et al. (2023).

\begin{figure}
\centerline{\includegraphics[width=9truecm]{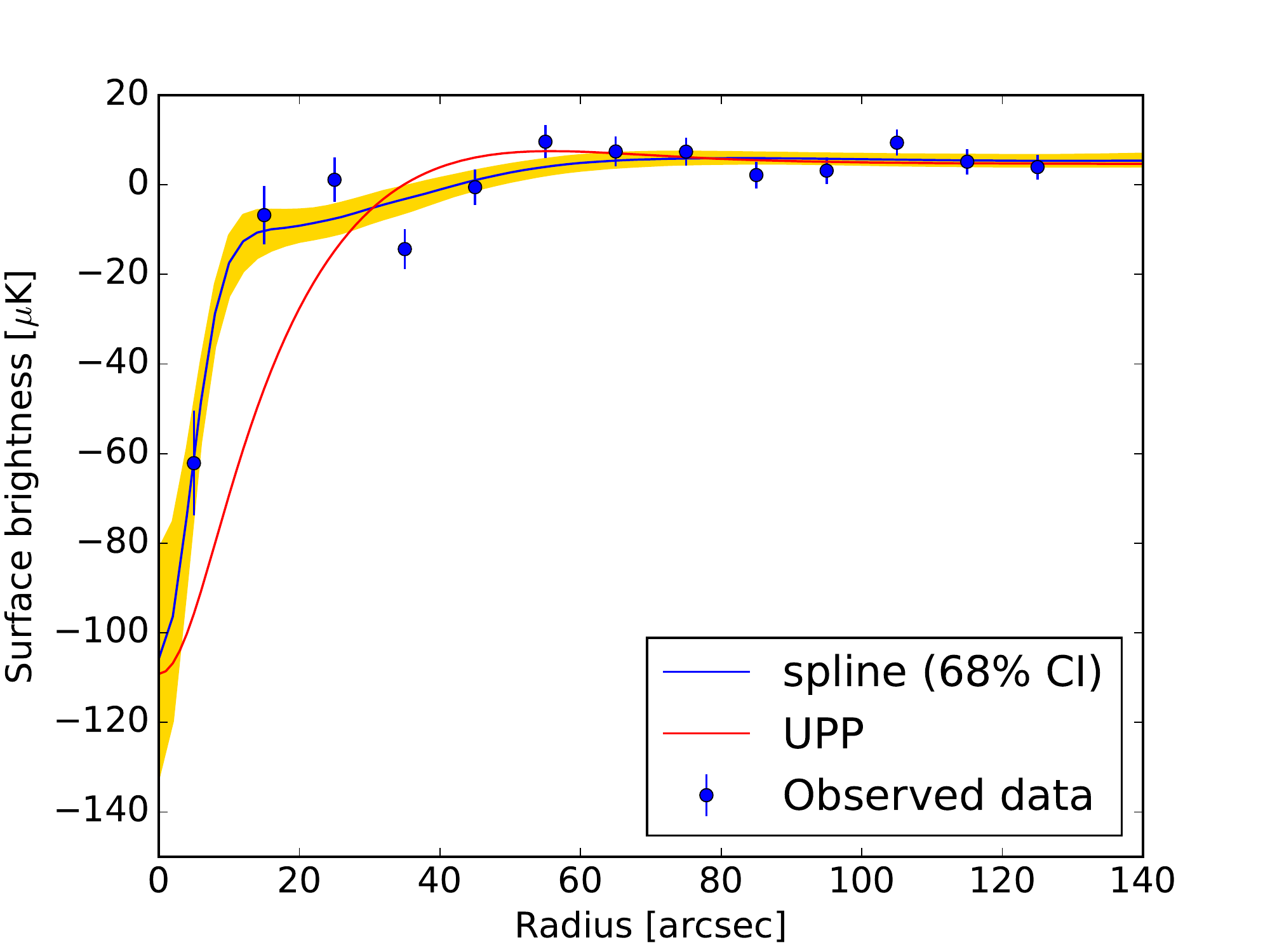}} 
\caption[h]{\JKCS SZ surface brightness profile (points with error bars) and 68\% uncertainties on the 
fitted model. The adopted spline for the pressure profile is able to well describe the observed data, 
whereas an universal pressure profile (UPP) with parameters fixed to Arnaud et al. (2010, red curve) does not.
}
\label{fig:surf_bright}
\end{figure}

\subsection{The depressed pressure profile}

Figure~\ref{fig:surf_bright} shows the cluster SZ radial surface brightness profile, 
extracted with 10 arcsec radial annuli centered on the SZ peak after flagging the point sources indicated in Fig.~\ref{fig:view}. The cluster is quite faint, with a peak of $\sim 60$ $\mu$K.
The brightness profile levels off to $\approx5$ $\mu$K, a non-zero level (that we name pedestal) induced by
the data reduction that forces the map to have zero mean.
 
The three dimensional pressure profile is derived fitting the SZ data with a modified version of \texttt{PreProFit} (Castagna \& Andreon 2019), accounting for the transfer function, point
spread function (beam), and pedestal level, following Andreon et al. (2021). 
The conversion from $K$ to Compton $y$ is derived in Andreon et al. (2021) and, since it is (minimally) $T$-dependent we adopt the 
observed temperature $kT=7.6$ keV (Andreon et al. 2009). 

Following Andreon et al. (2021), we adopt a pressure profile given by a cubic spline in log-log space with knots at radii of $r=5,15,30,$ and $60$ arcsec.  
By adopting a cubic spline we allow the shape of the pressure profile to vary almost arbitrarily
while keeping it continuous and doubly-differentiable.
By defining the spline in log quantities (log pressure vs log radius), we naturally exclude non-physical (negative) values
of pressure and radius and we can approximate a large variety of profiles, and their
derivatives, including the commonly parameterised pressure profile as in Nagai et al. (2007). 
Our model then has 5 variables:  the pressures at the four radii, $P_0,P_1,P_2,$ and $P_3$, and the pedestal value of the SZ surface brightness.  Our analysis assumes spherical symmetry,
uniform priors, zeroed for unphysical values (e.g., pressure cannot be negative).
In particular, since the total mass of the cluster is finite, the logarithmic slope
of the pressure should be steeper than $-4$ at large radii (Romero et al. 2018). We therefore
adopted a logarithmic slope $<-2$ at $r=1$ Mpc as prior. This prior
could alternatively be expressed as a maximal value for the pedestal level.
The imposition of this prior is  necessary primarily to distinguish a radially-flat background from a radially-flat
cluster signal.

\begin{figure}
\centerline{\includegraphics[width=9truecm]{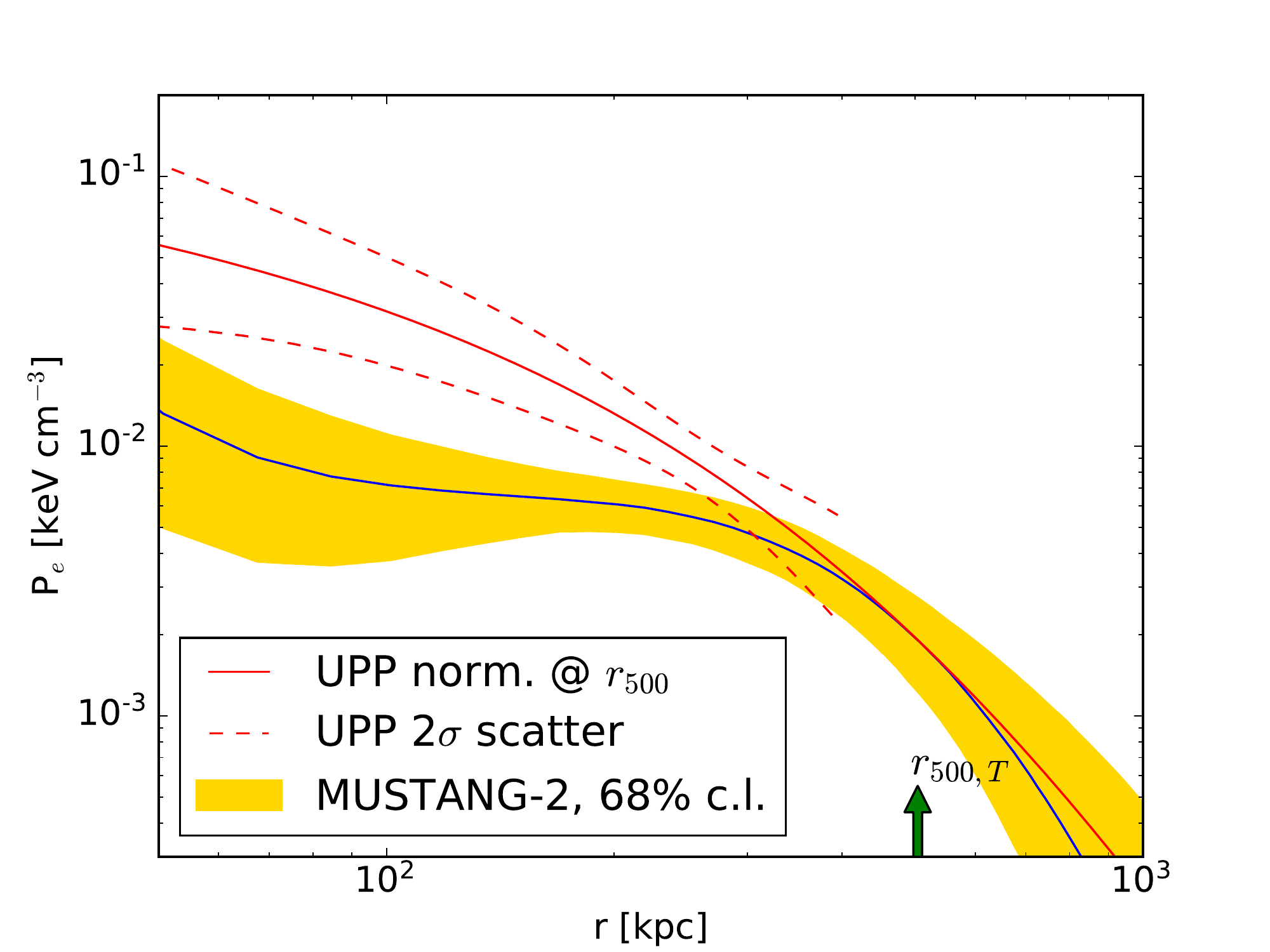}}
\caption[h]{Observed MUSTANG-2 \JKCS electron pressure profile with 68\% uncertainties (solid curve and shading) 
versus an universal pressure profile (UPP) with same pressure at $r_{500}$. 
Note that the displayed radial range includes radii poorly
sampled by MUSTANG-2 ($r\gtrsim 1$ arcmin, about 500 kpc at the cluster redshift). The Arnaud et al. (2010) scatter around the universal pressure profile is also indicated.}
\label{fig:pressure}
\end{figure}

\begin{figure}
\centerline{\includegraphics[width=9truecm]{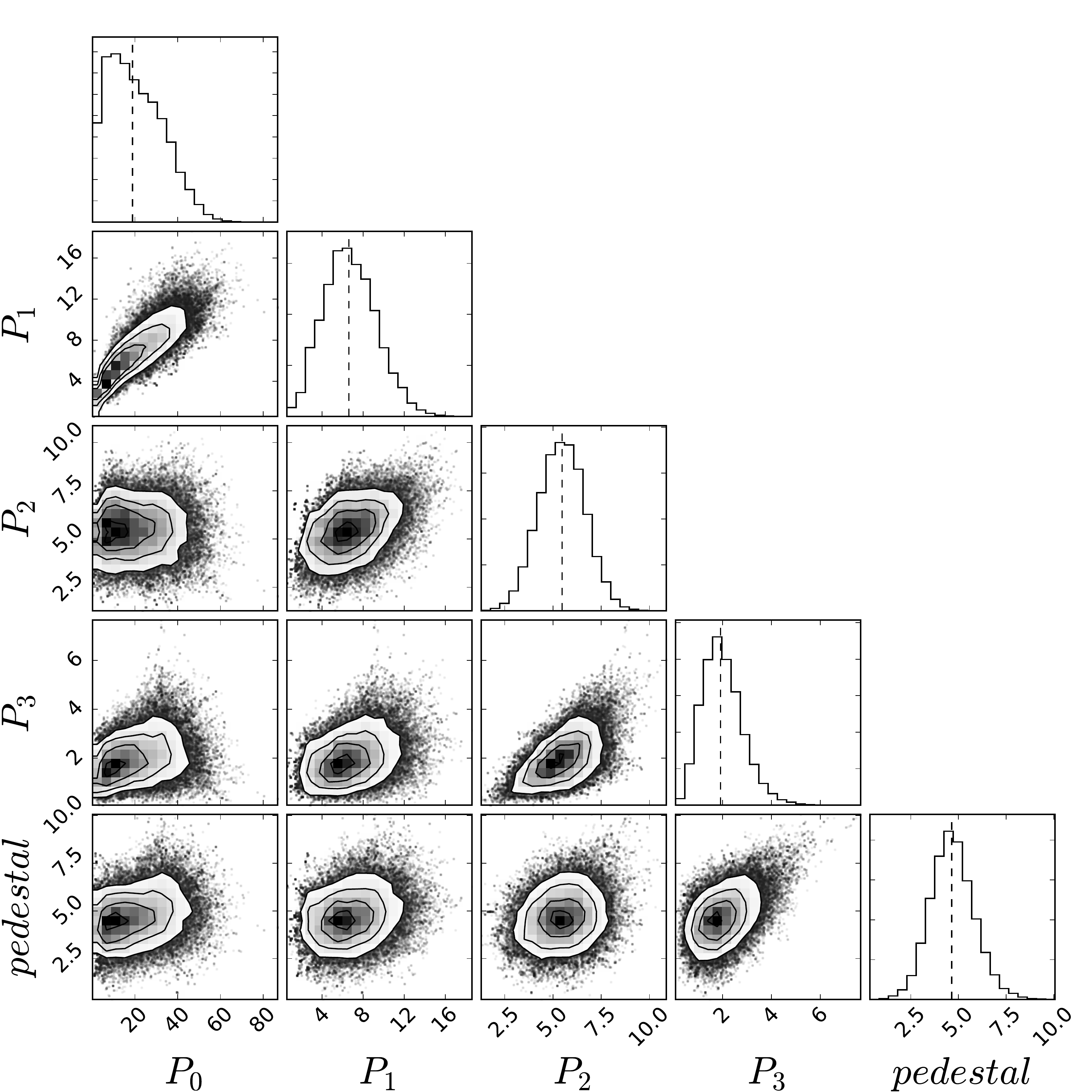}}
\caption[h]{Joint and marginal probability contours for our SZ fit. Units: eV cm$^{-3}$ for
pressure parameters and $\mu K$ for pedestal. 
}
\label{fig:cornerplot_sz}
\end{figure}

We find $P_i=19\pm13, 7\pm3, 5\pm 1, 1.9\pm0.7$ eV cm$^{-3}$, 
and a pedestal level of $5\pm1$ $\mu$K with covariances
illustrated in Figure~\ref{fig:cornerplot_sz}. We found a $\chi^2$ of $17$ for 8
degrees of freedom. There is a $\sim 2$\% probability to observe a larger $\chi^2$ assuming the model best fit.
The two points at $r\sim25$ and $35$ arcsec carry the largest contributions to the total
$\chi^2$, with the outer radius corresponding to local (negative) maxima in the SZ signal
at NE and SW in Fig.~\ref{fig:view} likely induced by the cluster merger.
Figure~\ref{fig:surf_bright} shows the best-fit model
on top of
the observed data. Figure~\ref{fig:pressure} shows the pressure profile as a
blue line with 68\% uncertainty as the shaded area.
This is the most distant cluster with a resolved SZ-based pressure profile, at only marginally larger
redshift than the previous record holder \IDCS (Andreon et al. 2021).

The \JKCS integrated spherical and cylindrical Componization parameter within $r_{500}$ are $7.7\pm1.8 \ 10^{-6}$ Mpc$^{2}$ and  $10\pm1 \ 10^{-6}$ Mpc$^{2}$, respectively. 
The \JKCS integrated cylindrical Componization parameter within 30 arcsec is $4.1 \ 10^{-6}$ Mpc$^{2}$, consistent with (below)  the 95\% upper limit of $6.8 \ 10^{-6}$ Mpc$^{2}$ reported in
Culverhouse et al. (2010). The latter is based on an inward extrapolation of the (lack of) signal measured on larger scales assuming  a 
radial profile for \JKCS close to the universal pressure profile, which however does not fit our data (Figs.~\ref{fig:surf_bright} and~\ref{fig:pressure}).

Fig.~\ref{fig:pressure} shows that \JKCS pressure profile is centrally-depressed compared to 
the universal pressure profile (UPP, Arnaud et al. 2010) being outside
the $2\sigma$ UPP range at several radii (larger significance requires a larger comparison sample).
In this comparison, we adopt Arnaud et al. (2010) best fit values, $r_{500}$ as derived from the X-ray temperature in
Andreon et al. (2014) and we fixed the  normalization to have a match at $r_{500}$. How well (badly) this
profile matches the observed data is shown in Fig.~\ref{fig:surf_bright} (red line). The shapes of the profile
do  not match: \JKCS has
a small-size central emission over a faint flat profile, unlike the universal pressure profile, that is spatially larger and slowly changing. 
Using the weak-lensing mass
instead of the one estimated from temperature 
does not alter our conclusion.
This is the third example of a cluster selected independently of the ICM content whose 
pressure profile turns out to be clearly depressed (with CL2015, Andreon et al. 2019 and \IDCS, Andreon et al. 2021), 
confirming previous claims that were either indirect (Di Mascolo et al. 2019) or of lower significance (Dicker et al. 2020). However, \JKCS and \IDCS are manifestly observed during their ICM assembly, and perhaps for this
reason their ICMs have not yet reach their final configuration, unlike CL2015. 

We have investigated whether the observed deviations of the SZ emission from a circular symmetric universal pressure profile centered on the X-ray center might be due to point sources. However, we find this unlikely because an universal pressure profile has an extent much wider than
the MUSTANG-2 PSF. In fact, at least four sources would be required to match the extent, but: a) we expect about 0.2 sources on average
from number counts; b) there are no sources detected at 90 GHz; and c) there are
no sources detected at 1.4 or 855 GHz, which instead would be expected for 90 GHz sources with a non-peculiar spectral energy distributions. In detail,
only focusing to the inner 30 arcsec, we need at least four sources
all with similar fluxes (70-100 $\mu$Jy), appropriately spaced apart (about 17 arcsec from each other), and with almost unique spectral energy distributions in order not to be detected in the available VLA data at 1.4 GHz (Bondi et al. 2003) and in SPIRE 855 GHz data (Schulz et al. 2017). In particular, a 70 $\mu$Jy source at 90 GHz would be detected at 1.4 GHz if its spectral slope is steeper than $-0.2$ and at 855 GHz if its spectral emission matches a gray body of 40 K (adopted to estimate the contamination in e.g. Sayers et al. 2013). 
This large density of ad-hoc sources in a small solid angle (with p-value probability $5.7 \ 10^{-5}$ based on the 0.2 sources observed on average in a 30 arcsec aperture by Zavala et al. 2021) and the required characteristics would still not be able to mimic the profile at larger scales nor to make the SZ emission circularly symmetric and co-centered with the X-ray. For the sake of completeness, we also consider the case in which the
matching to flux and extent of the universal pressure profile is achieved by contamination by many faint sources. To explore this case, we randomly draw distributions of fluxes from the Zavala et al. (2021) number counts (extrapolated down to 10 $\mu$Jy) within a solid angle with 30 arcsec radius and we count how many times we obtain a total net flux 
larger than 320 $\mu$Jy 
(to match the missing flux of the universal pressure profile within the considered solid angle) with individual fluxes in the simulated cluster aperture fainter than 70 $\mu$Jy (to avoid to be detected at 1.4, 90 or 855 GHz). This very rarely happens in our simulations, about twice every $10^5$ times. This computation neglects the fact that we need a peculiar spatial distribution of sources: a higher flux at the center of the aperture (where the SZ signal should be made larger) and an appropriate azimuthal flux distribution to achieve
a spherically symmetric profile centered on the X-ray center starting from an observed SZ distribution with a peak 26 arcsec away from it. This configuration is highly unlikely.
To summarize, we could not find a reasonable configuration of points sources that could explain the observed mismatch between the X-ray and SZ centers, the asymmetric SZ emission, and the excess at 35 arcsec, given the much larger extent of the cluster emission compared to the PSF. 
We therefore conclude that  point sources are extremely unlikely to alter our results.

\begin{figure}
\centerline{\includegraphics[width=9truecm]{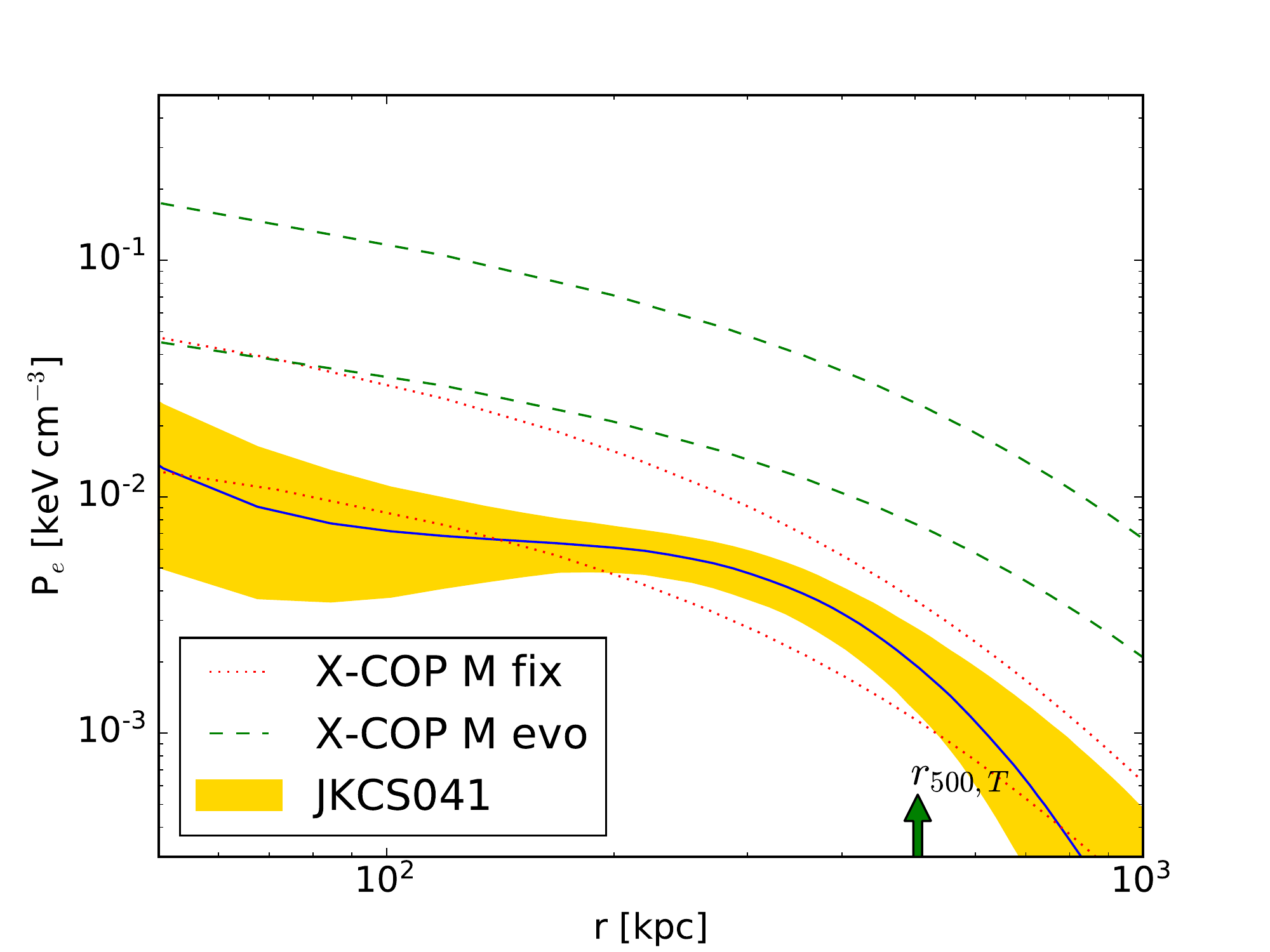}}
\caption[h]{\JKCS pressure profile (mean value and 68 per cent uncertainties,
blue line, and yellow shading) is compared to present day clusters with the same $M_{500}$ 
(the red dotted lines mark the $2\sigma$ range) and to
its present-day expect descendants (the dashed green lines mark the $2\sigma$ range inclusive of the
uncertainty on the mass accretion history). Compared to its present-day descendant, the pressure is low at
all radii. To become a present-day cluster, the pressure must increase by 0.7 dex. Observed values at $r>500$ kpc (about 1 arcmin)
are very uncertain because these scales are heavily filtered.
}
\label{fig:compXCOP}
\end{figure}

\section{Plausible evolution}

Fig.~\ref{fig:compXCOP} compares the \JKCS pressure profile (blue solid line) to a library of profiles derived for clusters in the nearby Universe (corridors).
The red dotted corridor indicates the $\pm 2\sigma$ range as derived by X-COP (Ghirardini et al. 2019) for clusters with \JKCS $M_{500}$
mass.  Similarly to Fig.~\ref{fig:pressure}, 
\JKCS has a low pressure for its mass in the inner 200 kpc. This comparison, at fixed mass,
is in our opinion misleading to study evolution: 
the  mass of \JKCS cannot stay constant for about 10 Gy, not even if the cluster acquires zero mass, because the integration radius, corresponding
to the overdensity $\Delta$, becomes 
larger at low redshift as a result of the
decrease of the mean density of the Universe (see Andreon et al. 2021 for discussion). 
To study evolution it is preferable to compare ancestors and descendants
(Andreon \& Ettori 1999). 
By selecting the 22 most massive 
clusters in the closest snapshot ($z = 1.71$) and their $z = 0$ descendants
in the Magneticum
simulation (Dolag et al. 2016, Ragagnin et al. 2017, Hirschmann et al. 2015),
we find that by $z = 0$ the mass of \JKCS will plausibly have grown by $0.60 \pm 0.16$ dex. This computation
actually uses clusters of slightly less mass than \JKCS ($0.64<M_{500}/M_\odot<1.4$ vs $2.2 \ 10^{14}$) because of the limited volume of the
simulation. The 
comparison of \JKCS to its descendants   
is shown by the green dashed corridor, marked by
the $\pm 2\sigma$ range. This, wider, corridor also accounts for 
the uncertainty on the mass growth. Compared to the library of possible descendants, the \JKCS pressure profile
is low at all radii by about 0.7 dex.

\begin{figure}
\centerline{\includegraphics[width=9truecm]{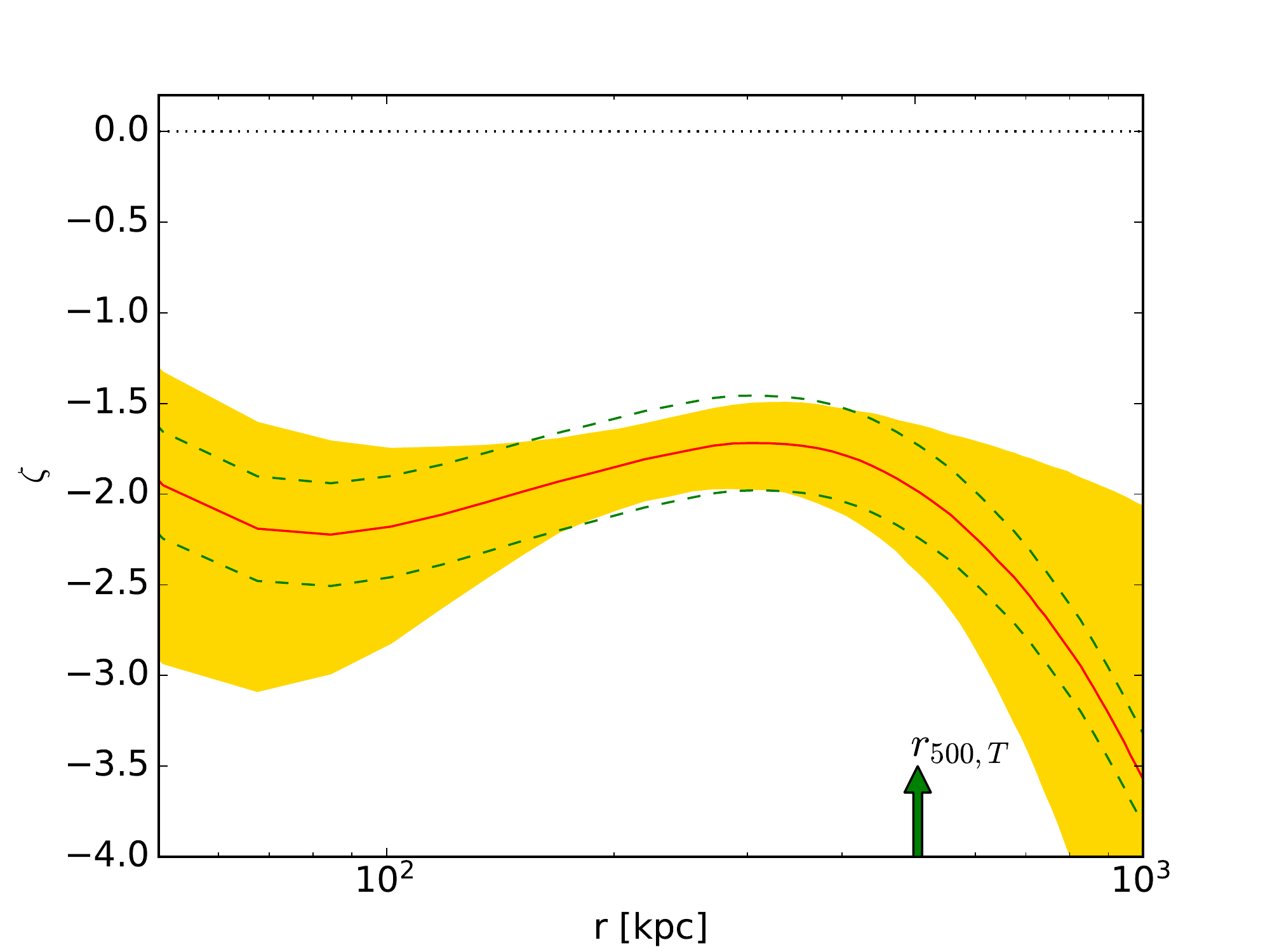}}
\caption[h]{Pressure evolutionary rate $\zeta$ vs. $r$ (eq.~1). 
The redshift evolution function $\zeta(r)$ of the \JKCS pressure
profile (mean value and 68 per cent uncertainties, red line and yellow
shading) is shown when compared to its present-day expected descendants. The dashed
corridor marks twice the X-COP scatter about the average and accounts for the uncertain mass accretion history. 
Negative values indicate a lower pressure in the past, while the no-evolutionary case correspond
to $\zeta = 0$. The plot quantifies the evolutionary rate already visible in
Fig.~\ref{fig:compXCOP}. The observed rate at $r>500$ kpc
is very uncertain because it is derived from data heavily filtered on these scales.
} 
\label{fig:evo}
\end{figure}

\begin{figure}
\centerline{\includegraphics[width=9truecm]{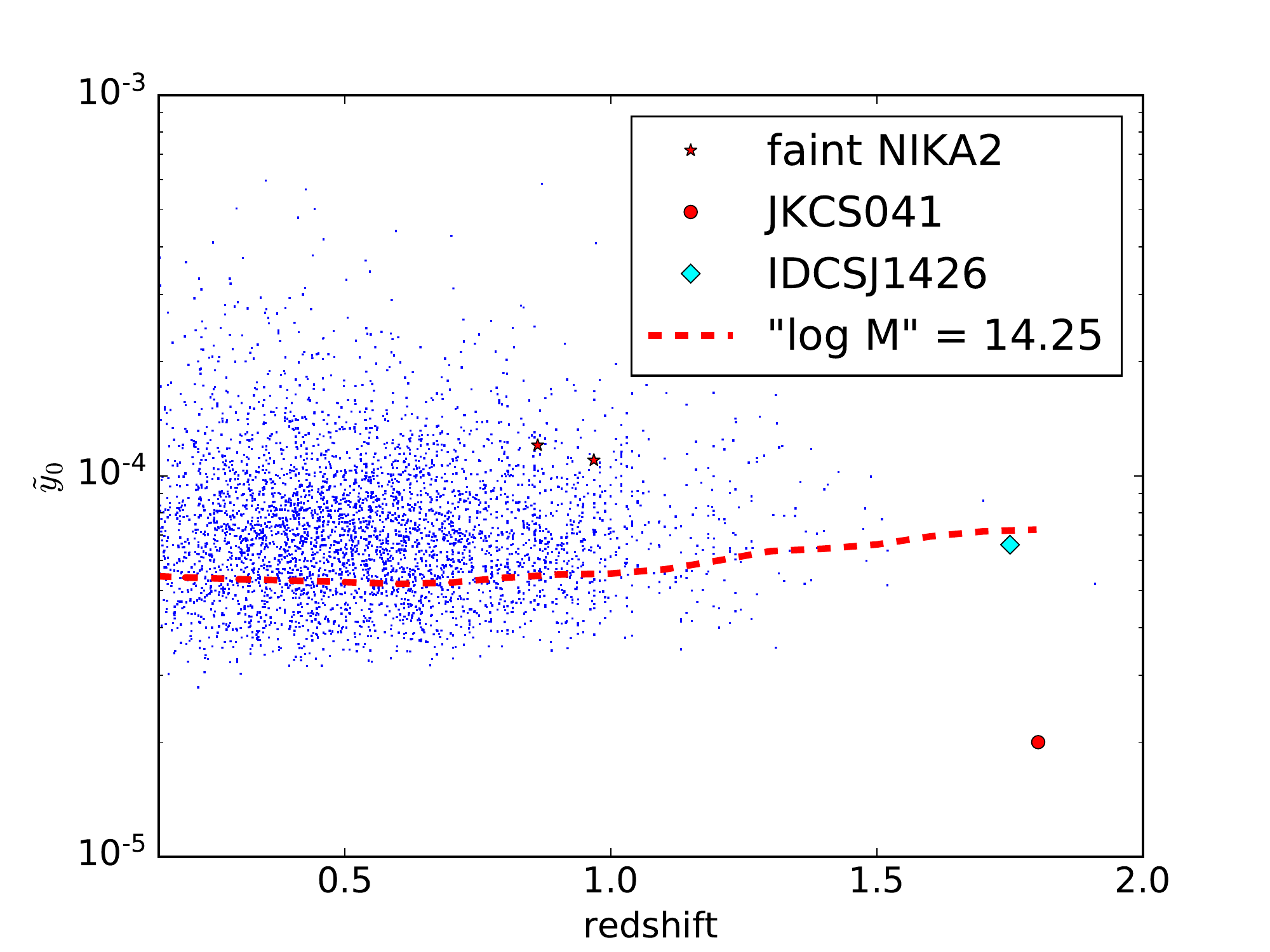}}
\caption[h]{Central comptonization parameter $\tilde{y}_0$ of various samples. ACT clusters in Hilton et al. (2021) are indicated with small points, \JKCS and \IDCS are marked with a red circle and  
cyan diamond, respectively, whereas two clusters denoted as very weak in the paper presenting their SZ signal 
are identified by stars at $z\sim0.9$. 
\JKCS is fainter by a factor of a few than these clusters.
The $\log M_{500}=14.25$ M$_\odot$ line is indicated, corresponding to the 10\% ACT completeness.
This plot is
the common mass-redshift plot except that here we use the observable $\tilde{y}_0$ instead of
identifying it with mass. Using mass produces a slight vertical shift of each point such that the red dashed line becomes straight and flat. 
}
\label{fig:y0_z}
\end{figure}

To quantify the amount of evolution, following Andreon et al. (2021) we compute the evolutionary rate $\zeta$, which is the exponent of $E_z$ in:
\begin{equation}
P(r,z,M_z)=P(r,z=0,M_{z=0}) E_z^\zeta
\end{equation}
where $E_z=H(z)/H_0$ and $M_{z=0}$ is the $z=0$ mass of the plausible descendant of the cluster with mass $M_z$ at
redshift $z$.  
Unlike other authors, we compare same radii in kpc, not radii scaled by $r_{500}$ because the cluster
content does not expand with the Universe (see Andreon et al. 2021 for details). As local, comparison, sample, we use the pressure profile derived from the X-COP sample (Ghirardini et al. 2019).
Fig.~\ref{fig:evo} shows the derived evolutionary rate and quantifies what is already visible in
Fig.~\ref{fig:compXCOP}: pressure is lacking at high $z$ and must increase with a rate $\zeta\approx-2$ at most
radii. The figure shows the uncertainty due to measurement errors (yellow shading) and that from the combination of the one coming from the
variety of profiles at a fixed mass at $z=0$ and of the scatter in accretion histories (dashed green corridor, marking
the $\pm 2\sigma$ range). The large pressure change implied in this comparison indicates that \JKCS is in a very active phase,
as also shown by the SZ-X offset. Unlike \IDCS, the inner 200 kpc display a large evolutionary rate $\zeta$, indicating that
even the central part of \JKCS is far from the likely final status, in agreement with the SZ-X offset.

One caveat applies: the $z=0$ clusters that we use as a local sample are ICM selected, either in X-ray
(Fig.~\ref{fig:pressure}) or in SZ (Fig~\ref{fig:compXCOP}), 
whereas \JKCS is selected independently of its ICM content because it was
discovered in near-infrared (and then followed-up with Chandra and MUSTANG-2). ICM-selected clusters have, by selection, pressure profiles
biased high and with reduced scatter because clusters with low pressure are underrepresented, or absent,
in those samples (Andreon et al. 2016, 2019). A cluster with the \JKCS Compton parameter $Y_{SZ}$, or central Compton parameter $\tilde{y}_0$, is undetectable
in the ACT (Hilton et al. 2021) or SPT (Bocquet et al. 2019) as discussed in Sec.~\ref{sec:faint}, and also with Planck 
(Planck collab. 2016) when $z\gtrsim0.02$. Therefore, if sizeable samples
of clusters with depressed profiles exist in the nearby Universe, the range seen in REXCESS or X-COP is reduced and
biased, and our estimate of evolution biased high. We have demonstrated that at least one such cluster exists, CL2015 (Andreon et al. 2019).
The same caveat applies to estimates of evolution by other authors (e.g. Pratt et al. 2022) that furthermore
mix evolution with pseudo-evolution by scaling radii (see Andreon et al. 2021 for details and comments on
earlier estimates).

The observed $\zeta\approx-1.5$ evolutionary rate is similar to values seen in the 22 clusters in the Magneticum simulation at $r>200$ kpc
but smaller at smaller radii (rates for individual simulated clusters are shown in Andreon et al. 2021 Fig.~16). In particular, 
in simulations inner radii are already pressured (i.e., $\zeta \approx 0$) and, if any, are
over-pressured at high $z$ (i.e., $\zeta > 0$) compared to present day descendants. On the other hand, a similar
profile might be missing in the simulation simply because we  have only 22 simulated clusters (of slightly lower mass) and major head-on mergers of massive sub-clusters, as \JKCS is, are rare (also in samples of one hundred, Wik et a 2008).

\subsection{The faintness of \JKCS}
\label{sec:faint}

\begin{figure}
\centerline{\includegraphics[width=9truecm]{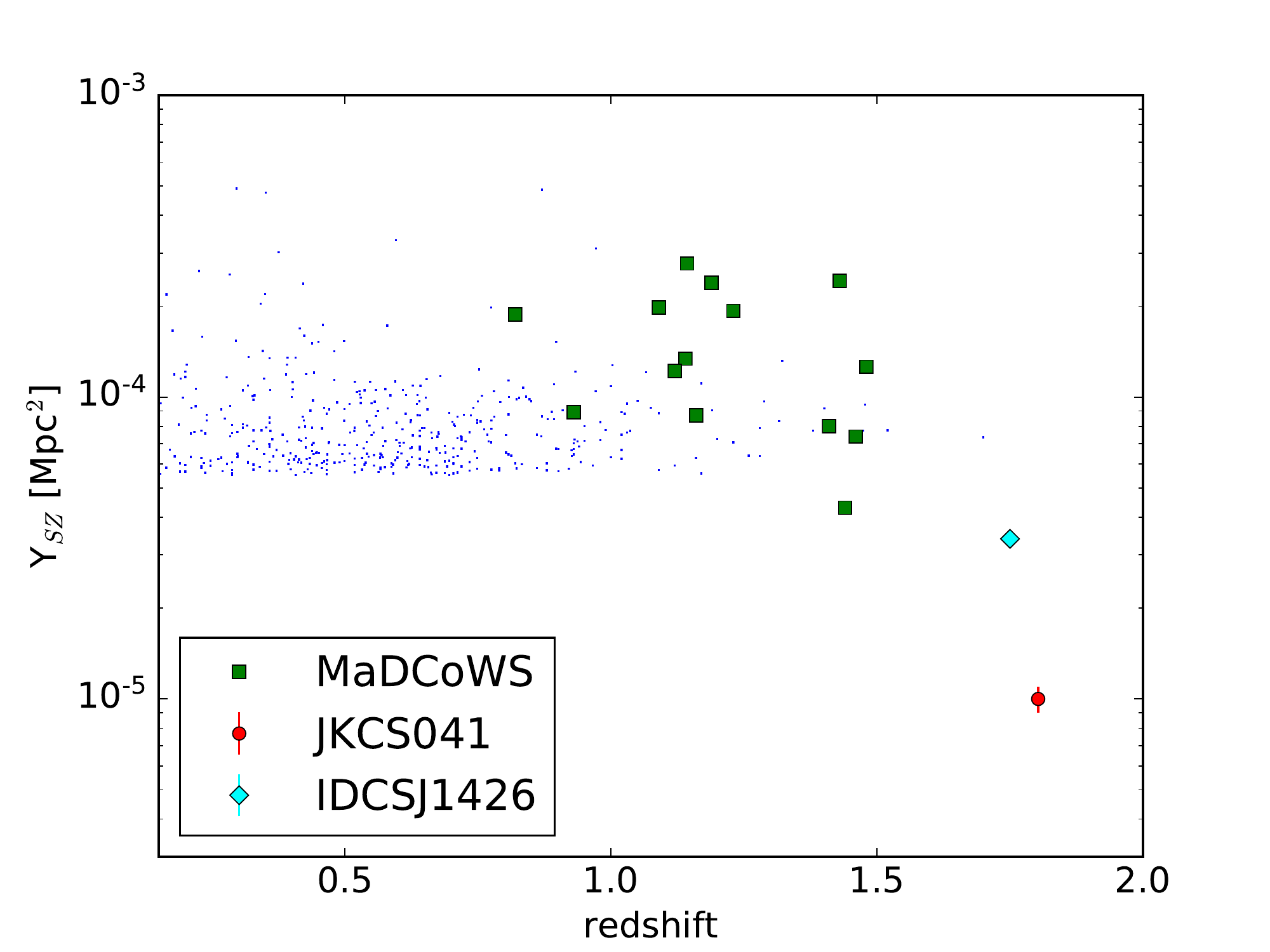}}
\caption[h]{Cylindrical integrated Compton parameter $Y_{SZ}$ within $r_{500}$ for SPT clusters in Bocquet et al. (2020, small points), \JKCS (red circle),  
\IDCS (cyan diamond), and MaDCoWS clusters (squares). \JKCS is fainter by a factor of a few than the faintest SPT and MaDCoWS clusters. This plot is
the common mass-redshift plot except that here we use the observable $Y_{SZ}$ instead of
identifying it with mass.}
\label{fig:y_SPT_z}
\end{figure}

Compared to other clusters with SZ observations, \JKCS is extremely faint, as illustrated in Fig.~\ref{fig:y0_z} comparing
its central central Compton value $\tilde{y}_0$ and literature values. ACT values (small points) are from 
from Hilton et al. (2021), while the \JKCS (red circle) and 
\IDCS (cyan square) $\tilde{y}_0$ values are derived using MUSTANG-2 data following the ACT definition (Hilton et al. 2021): we took the central value of the $y$ map,
derived by Abel (line of sight) projection of the pressure profile, and
filtered with the convolution of an universal pressure profile with scale 2.4 arcmin and the ACT beam. For \IDCS we used MUSTANG-2
data in Andreon et al. (2021). We found $\tilde{y}_0=2.0 \times 10^{-5}$ and $6.6 \times 10^{-5}$ for \JKCS and \IDCS, respectively.
\JKCS is
$\sim 4$ times fainter than the 10\% ACT completeness value, indicated by a dashed line. As a way of comparison, a couple of clusters, denoted as  very faint in the papers presenting them 
(K{\'e}ruzor{\'e} et al. 2020 and 
Ricci et al. 2020), are indicated
by stars (at $z\sim0.9$) and they are almost 10 times brighter than \JKCS .  Fig.~\ref{fig:y_SPT_z} repeats
the comparison using SPT clusters from Bocquet et al. (2019) and 
Massive and Distant
Clusters of WISE Survey (MaDCoWS) with a $\geq 5\sigma$ SZ detection
from Dicker et al. (2020). The faintest of these clusters are brighter
by a factor of a few compared to \JKCS . \JKCS is a really faint cluster and indeed its surface brightness profile is noisier than
that of \IDCS in spite of a much longer integration time at the same facility (5 vs 28 h).
This plot is
the common mass-redshift plot except that here we use the observable $Y_{SZ}$ instead of
identifying it with mass which removes any ambiguity.

\subsection{The SZ mass-mass plot}

\begin{figure}
\centerline{\includegraphics[width=9truecm]{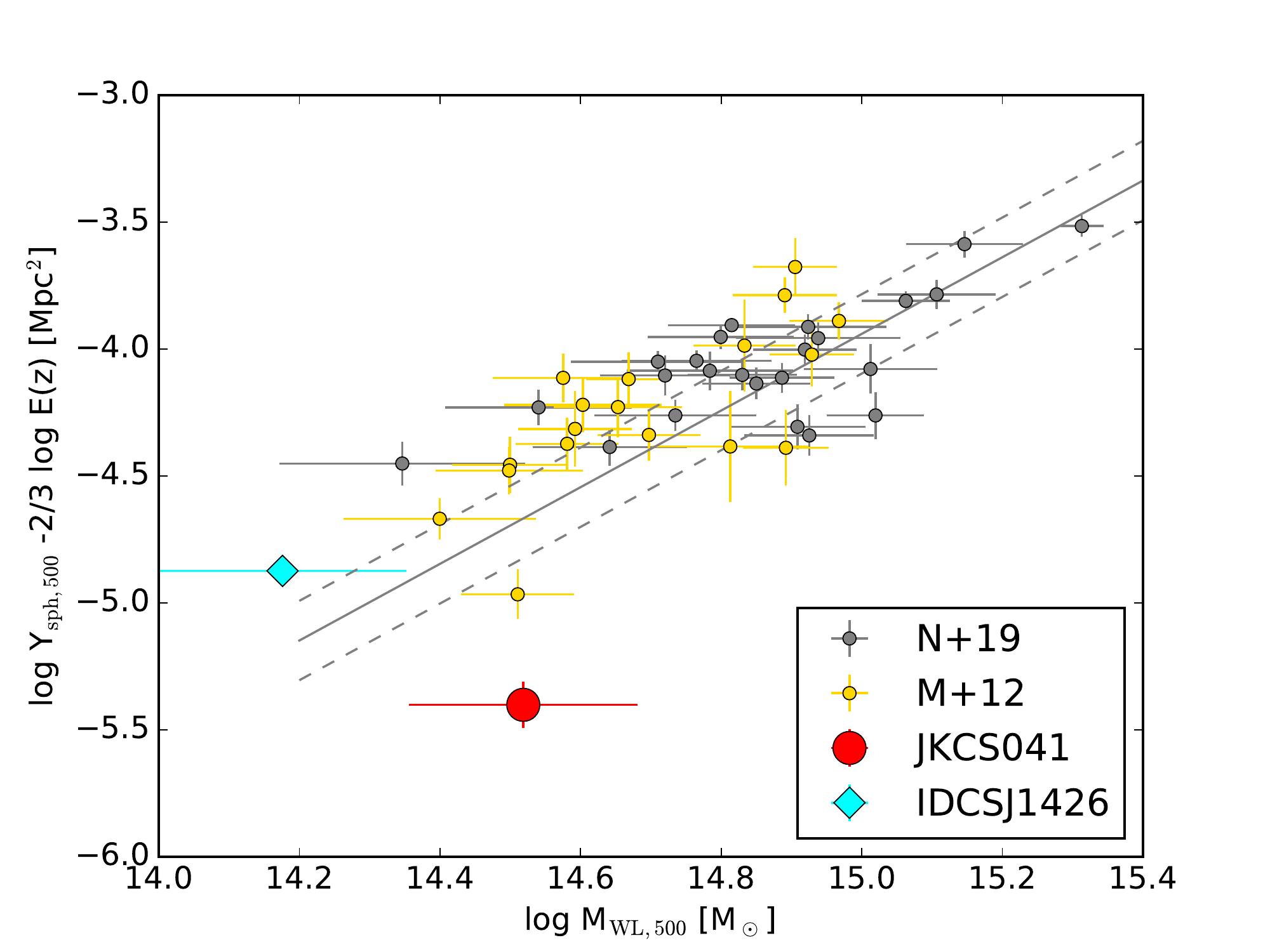}}
\caption[h]{Spherical $Y_{\rm SZ}-M_{\rm WL}$ scaling (aka the SZ mass vs mass plot) for JKCS\,041, \IDCS and clusters at $z<0.55$ (from Marrone et al., 2012 and Nagarajan et al., 2019).
The solid line indicates the fit to the latter sample 
while the dashed corridor is the mean model plus and minus the
intrinsic scatter (as computed by Nagarajan et al. 2019). \JKCS is 0.75 dex below the mean relation, and 
has a 0.5 dex fainter SZ signal than the lower-mass similar-redshift \IDCS cluster, casting doubts on the identification of mass with $Y$ assumed in many past papers.
}
\label{fig:wl}
\end{figure}

Fig.~\ref{fig:wl} shows the spherical Compton $Y$ within $r_{500}$ corrected for the self-similar
evolution vs. weak lensing mass for \JKCS, \IDCS and the ICM-selected $z<0.55$ clusters in Marrone et al. (2012) and Nagarajan et al. (2019). 
\JKCS mass is about 0.35 dex larger than the mass of \IDCS, yet the SZ signal is 0.5 dex lower, i.e. there is a sizeable
difference between the mass and the Compton parameter for these two clusters at $z\sim2$. Some authors might refer to it as a difference between between SZ mass and mass. This large difference for just two high redshift clusters alone is enough to cast doubts on the identification of mass with the Compton parameter
assumed in many past papers. Indirect evidence of this large scatter is seen in richness vs. $Y_{SZ}$ plane, with $Y_{SZ}$ identified with mass, of $z\approx 1$ clusters in Di Mascolo et al. (2020).
Under the assumption that the Compton parameter measures mass, Brodwin et al. (2016) concluded that  \JKCS has a much lower mass than \IDCS , in spite 
of its much larger richness of \JKCS (Andreon 2013) and of the larger estimate of the 
 weak lensing mass of \JKCS.

Fig.~\ref{fig:wl} shows that \JKCS is 0.75 dex below the mean mass vs Compton $Y$ relation of the ICM-selected $z<0.5$ clusters from Nagarajan et al. (2019).
The low \JKCS Compton parameter for its mass is in line with expectations for merging clusters, observed in simulations to scatter below the mean relation (Wik et al. 2008; Krause et al. 2012).  
Indeed, according to simulations, merger events cause clusters to scatter downward in this plane because the $Y$ increase is delayed compared the mass growth and because the final $Y$ is lower than that for a cluster of the same final mass (Wik et al. 2008; Krause et al, 2012). According to simulations, the same downscatter would happen to many
high redshift clusters, because at early epochs the fraction of disturbed or merging clusters is high (Krause et al, 2012). Therefore, we expect a large scatter between SZ mass and true mass at high redshift, with SZ surveys becoming more and more biased against merging cluster as redshift increases. According to simulations,
\JKCS final equilibrium $Y$ value will be larger 
by  $\sim 0.2$ dex (Wik et al. 2008), 
insufficient to significantly reduce the SZ mass difference between the two high redshift clusters.
As for other comparisons with lower redshift samples (Figs.~\ref{fig:pressure}, \ref{fig:compXCOP}, \ref{fig:evo}, \ref{fig:y0_z}, and \ref{fig:y_SPT_z})
the relative location of \JKCS with respect to the mean relation in Fig.~\ref{fig:wl}) is affected by affected
by a hard-to-quantify bias in current low redshift ICM-selected
samples.

\begin{figure}
\centerline{\includegraphics[width=9truecm]{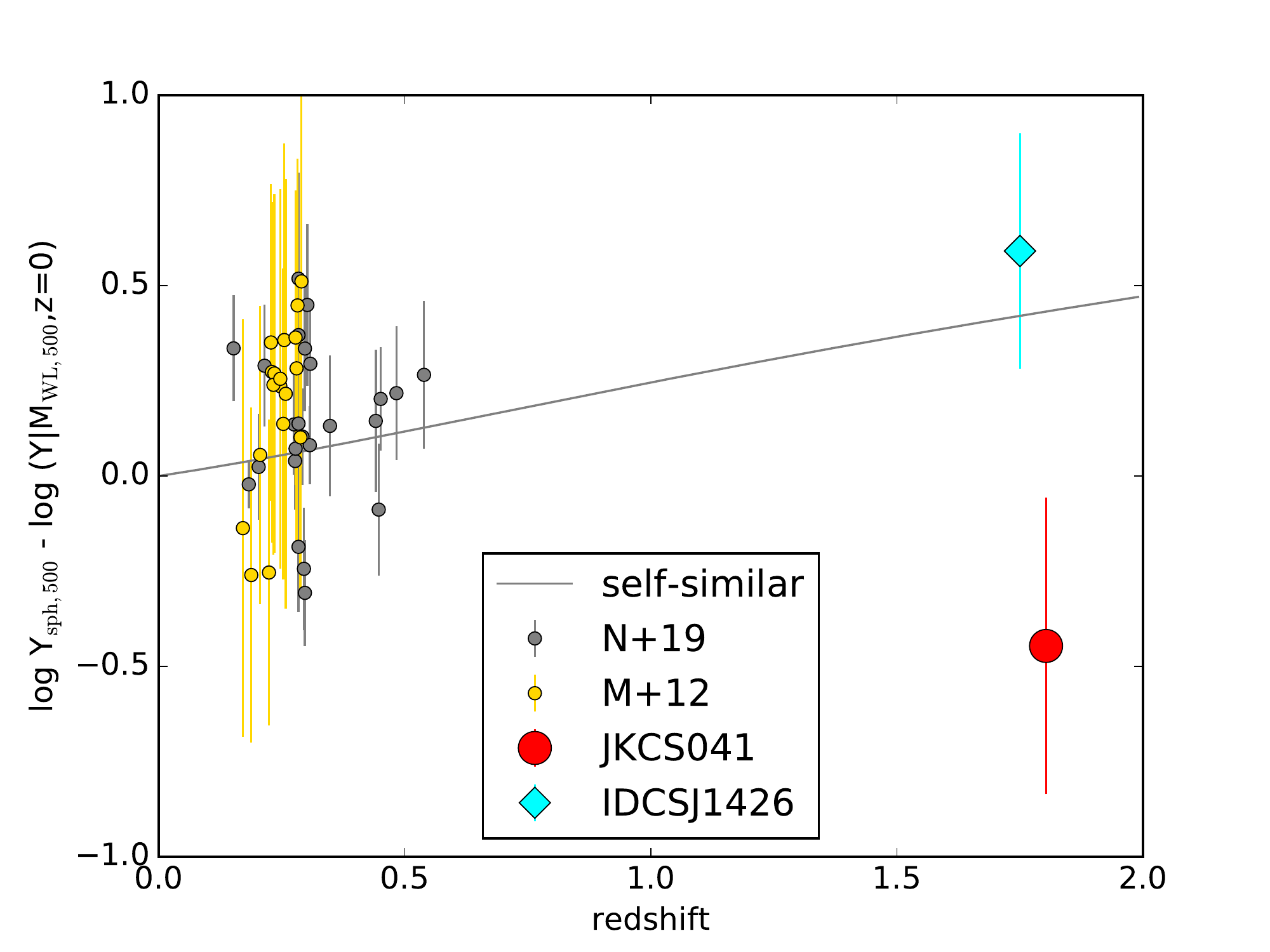}}
\caption[h]{The uncertain evolution of the Compton $Y$ vs mass relation. 
Distance from the $z=0$ mean relation of spherical $Y_{\rm SZ}$ vs redshift for
JKCS\,041, \IDCS and clusters at $z<0.55$ (from Marrone et al. 2012 and Nagarajan et al. 2019).
The solid line indicates the $E^{2/3}_z$ self-similar evolution. \JKCS is 0.9 dex below it. 
}
\label{fig:evoYz}
\end{figure}

Fig.~\ref{fig:evoYz} shows the same data as previous figure, but with a focus on evolution. The ordinate is the Compton Y distance (at fixed mass) from the $z=0$ mean relation. The line is the $E^{2/3}_z$ self-similar  evolution. Errorbars account for, and are dominated by, mass errors. The figure emphasizes the limited knowledge of evolution of this scaling relation at the cosmic noon due to the dearth of at $z\sim 2$ data.
\JKCS is 0.9 dex below the self-similar prediction, although with a large error. Also in this plot, the relative 
\JKCS location compared to other samples is
partially amplified by a hard-to-quantify bias in current low redshift ICM-selected samples. Furthermore,
we note 
that the observed mass at the cluster redshift is used to estimate the predicted $Y$ using the $z=0$ relation ignoring mass growth and pseudo-evolution between the two considered redshifts (see Andreon et al. 2021).

\subsection{Summary}

We present a 28h long Sunyaev-Zel'dovich (SZ) observation of the $z=1.803$
\JKCS cluster taken with MUSTANG-2 at the Green Bank Telescope. This very long exposure allows 
us to detect \JKCS, even if the cluster is intrinsically extremely weak  compared to other SZ-detected clusters.

We found that the SZ peak is offset from the X-ray center by about 220 kpc in the  direction of the brightest cluster galaxy, which we interpret as due to the cluster being observed just after a major merger, 
as a part of the emergence of the first massive
clusters from the  cosmic web at early times.

We found that \JKCS has a low central pressure compared to local ICM-selected clusters and a low Compton Y for its mass, likely because the cluster is
still in the process of assembly, but also amplified by a hard-to-quantify bias in current low redshift ICM-selected samples. 
These observations provide us with the measurement of the most distant resolved pressure profile.
Comparison with a library of plausibly descendants shows that \JKCS pressure profile will likely increase by about 0.7 dex in the next 10 Gyr at all radii.

\JKCS has an SZ mass $0.5$ dex lower than the lower-mass $z=1.75$ 
\IDCS cluster, exemplifying at $z\sim2$ how much different true mass and SZ mass can be.

\section*{Acknowledgements}
SA thanks Sharon Felix, James Jee, Jinhyub Kim, and Lindsay King for useful discussion and for sharing results in advance of 
publication.
The Green Bank Observatory is a facility of the National Science Foundation operated under cooperative agreement by Associated Universities, Inc. 
MUSTANG-2 data was taken under the project ID AGBT18B-111, AGBT19B-003, AGBT20B-020.
CS was supported in part by NASA Chandra grant G01-22120X and XMM Grant 80NSSC22K0857.

\section*{Data Availability}

Raw Chandra data are public available in the Chandra archive, obs ID 15168 and 16321. SZ derived products (beam, transfer function, and map) are available at https://doi.org/10.7910/DVN/AYSGA3. 

{}

\bsp	
\label{lastpage}
\end{document}